\documentclass{emulateapj}
 \usepackage{psfrag}
 \usepackage{rotating}
 \usepackage{color}
 
\shorttitle{Interferometry \& the Stellar Mass Index}
\shortauthors{Neilson~et~al.}

\begin{document}

\title{Stellar atmospheres, atmospheric extension and fundamental parameters: weighing stars using the stellar mass index}

 \author{Hilding R.~Neilson\altaffilmark{1}}
 \altaffiltext{1}{Department of Astronomy \& Astrophysics, University of Toronto, 50 St.~George Street, Toronto, ON, M5S 3H4, Canada}
\email{neilson@astro.utoronto.ca}
\author{Fabien Baron\altaffilmark{2}} 
\author{Ryan Norris \altaffilmark{2}}
\author {Brian Kloppenborg\altaffilmark{2}}
\altaffiltext{2}{Center for High Angular Resolution Astronomy, Department of Physics and Astronomy, Georgia State University, P. O. Box 5060, Atlanta, GA 30302-5060, USA;}
\author{John B.~Lester\altaffilmark{1,3}}
\altaffiltext{3}{Department of Chemical \& Physical Sciences, University of Toronto Mississauga, Mississauga, ON L5L 1C6, Canada}

\begin{abstract}
One of the great challenges in understanding stars is measuring their masses. The best methods for measuring stellar masses include binary interaction, asteroseismology and stellar evolution models, but these methods are not ideal for red giant and supergiant stars. In this work, we propose a novel method for inferring stellar masses of evolved red giant and supergiant stars using interferometric and spectrophotometric observations combined
with spherical model stellar atmospheres to measure what we call the stellar mass index, defined as the ratio between the stellar radius and mass. The method is based on the correlation between different measurements of angular diameter, used as a proxy for atmospheric extension, and fundamental stellar parameters. For a given star, spectrophotometry measures the Rosseland angular diameter while interferometric observations generally probe a larger limb-darkened angular diameter. The ratio of these two angular diameters is proportional to the relative extension of the stellar atmosphere, which is strongly correlated to the star's effective temperature, radius and  mass. We show that these correlations are strong and can lead to precise measurements of stellar masses.
\end{abstract}
\section{Introduction}

Red giant and supergiant stars represent a crucial stage of stellar evolution where stars fuse elements heavier than hydrogen as they approach the end of their lives before they explode as core-collapse supernovae,  driving the next generation of star formation, or shed their envelopes on the evolutionary path to becoming white dwarfs.  These stars  have the largest radii of all stars and therefore, are ideal targets for optical interferometric observations \citep[\textit{e.g.}][]{Kloppenborg2015}.

Interferometric observations of red supergiant stars have provided insights into  center-to-limb intensity variation (CLIV), also known as limb darkening, convection, deviations from spherical symmetry, and circumstellar material \citep[][and others]{Haubois2009, Ohnaka2011}.  While these observations are powerful and insightful, there exist many challenges for interpreting these measurements and determining fundamental stellar parameters, such as mass. The measurement of stellar mass is especially problematic because there are only a few viable methods, some of which are sensitive to other processes, such as internal mixing and mass loss.

There are two primary tools for inferring stellar masses of single giant and supergiant stars: stellar spectroscopy and stellar evolution modelling. The latter method requires knowledge of many other fundamental parameters such as effective temperature and luminosity.  However, red giant and supergiant stars of different masses converge to similar effective temperatures near the Hayashi line on the Hertzsprung-Russell diagram \citep{Kippenhahn2012, Ekstrom2012}. For example, \cite{Dolan2016} determined the mass of Betelgeuse by fitting stellar evolution models to measurements of its radius, effective temperature and mass-loss rate and found a mass of about $20~M_\odot$. This value, however, depends on the treatment of mass loss, core convection, and internal mixing in the stellar evolution models, which in turn limits the robustness of the determination.

Similarly, stellar spectra are ideal for measuring effective temperatures, surface gravities and compositions 
\citep[\textit{e.g.}][]{Gray2005}.  For example, \citet{Lobel2001} measured $\log_{10} g \approx -0.5$ for Betelgeuse.  However, this value is also sensitive to turbulence and convective velocities \citep{Gray2012}, 
increasing the uncertainty of the final estimated mass.

\cite{Neilson2011a} and \cite{Lester2013} showed it is possible to infer a star's mass from the 
extension of the atmosphere by using measurements of the star's CLIV and stellar atmospheric models. \cite{Neilson2011b} measured a mass for Betelgeuse to be about 12 -- 16~$M_\odot$, much smaller than found using stellar evolution models.  While that method depended on limb-darkening laws \citep[\textit{e.g.}][]{Claret2000, Claret2011, Neilson2013a, Neilson2013b}, it suggested  the possibility of applying measurements of atmospheric extension for inferring stellar masses.

Optical interferometry measures both the angular diameter and stellar CLIV \citep{Davis2000}. \cite{Wittkowski2004} attempted to fit stellar atmosphere model intensity profiles directly to interferometric and spectrophotometric observations of a red giant star.  Although they were unable to distinguish between different stellar atmosphere models, they did measure precisely the limb-darkened and Rosseland angular diameters. Similar results were obtained for other red giant stars \citep{Wittkowski2006a, Wittkowski2006b}. A key issue was that these studies had to assume stellar parameters for the atmosphere models. But using a grid of model stellar atmospheres computed as described in \citet{Lester2008}, \cite{Neilson2008} were able to fit both spectrophotometric and interferometric observations used by \citet{Wittkowski2004,Wittkowski2006a,Wittkowski2006b} without assuming any stellar parameters.  The resulting models reliably reproduced the measured Rosseland and limb-darkened angular diameters.  

However, for both analyses stellar masses were inferred from stellar evolution tracks; fitting model stellar atmospheres did not constrain the masses of the stars. The combination of interferometric and spectrophotometric observations alone are currently insufficient for measuring stellar masses. There are not, as of yet, methods to measure masses directly from fitting model stellar atmospheres to observations with any significant precision.

The purpose of the present work is to develop a new method for inferring stellar fundamental parameters from interferometric and spectrophotometric observations using spherical model stellar atmospheres. This approach is based on the previous work of \cite{Neilson2012} who found that coefficients from a specific limb-darkening law are correlated with the atmospheric extension.  In this work, we completely abandon the concept of arbitrary limb-darkening laws with adjustable coefficients, which in practice are poorly constrained by measurements.  In the next section we develop proxies for atmospheric extension based on spectro-interferometric measurements, tying them to fundamental parameters, especially stellar masses.  In Section~3, we outline the spherical model stellar atmospheres used in this work and present the best-fit correlations in Section~4 along with a comparison to previous measurements.  We summarize our work in Section~5.

\section{Atmospheric Extension and the Stellar Mass Index} \label{sec:sec2}
\citet{Neilson2012} showed that for spherically symmetric model
stellar atmospheres the coefficients used to fit the CLIV depend on the amount of atmospheric extension,  denoted by the ratio of the stellar radius to mass, $R/M$ in solar units. The stellar radius is defined as the point in the star where the Rosseland optical depth is 2/3 \citep{Mihalas1978}. Here we define this ratio as the stellar mass index (SMI), 
\begin{equation} \label{eq:smi}
\mathrm{SMI} \equiv \frac{R_{\rm{Ross}}}{M}~(R_\odot/M_\odot).
\end{equation}
 The SMI is analogous to a person's body mass index, the ratio of a person's height and weight.  These  results suggest that a measurement of atmospheric extension can provide information about fundamental stellar parameters, potentially without requiring knowledge of distances.

Following the approach of \cite{Baschek1991} and \cite{Bessell1991}, \cite{Neilson2012} represented  the relative extension of a stellar atmosphere as 
\begin{equation} \label{eq:dr_r1}
\frac{\Delta R}{R_\mathrm{Ross}} \propto \frac{H_\mathrm{P}}{R_\mathrm{Ross}} \equiv \frac{kT_{\rm{eff}}/\mu m_\mathrm{H} g}{R_\mathrm{Ross}},
\end{equation}
 where $H_\mathrm{P}$ is the pressure scale height and where $\Delta R \equiv R_{\rm{LD}} - R_{\rm{Ross}}$ is the difference between $R_{\rm{LD}}$, the radius at the top of the atmosphere, and $R_{\rm{Ross}}$, the Rosseland radius. The variables $\mu m_H$ represents the mean atomic mass in the atmosphere, independent of mass or radius for a given atmospheric composition.  Using $g \propto M_\star/R_{\rm{Ross}}^2$  gives
 \begin{equation} \label{eq:dr_r2} 
\frac{\Delta R}{R_\mathrm{Ross}} 
\propto T_\mathrm{eff} \left (\frac{R_\mathrm{Ross}}{M_\star} \right )
= T_\mathrm{eff} \times \mathrm{SMI}.
\end{equation}
This suggests that the stellar mass is defined, along with the gravity, at the Rosseland radius and not at the limb-darkening radius.  But, because the density in the photosphere is orders of magnitude smaller than in the stellar envelope and core, the difference in stellar mass at the Rosseland and at the limb-darkening radius is negligible.

 The theoretical definition of $R_{\rm{LD}}$ involves the minimum optical depth of the photosphere and hydrostatic equilibrium: the boundary condition at the top of the photosphere implies that the pressure and density go to zero. Consequently the precise value of the radius at the top of the photosphere is model dependent: it depends on the hydrostatic equilibrium in the model, i.e., on the gravity of the stellar atmosphere. In practice, boundary conditions are imposed by positing an actual minimum optical depth in model atmosphere codes (see e.g., section~\ref{sec:sec4} for the adopted value in \textsc{Atlas/SAtlas}). Different reasonable choices for this minimum optical depth lead to slightly different values $R_{\rm{LD}}$; however this spread in values remains insignificant for our purpose, being at least two orders of magnitude below expected $\Delta R$.

The $\Delta R/R_\mathrm{Ross}$ defined above can be measured empirically employing a combination of spectrophotometry and interferometry \citep[\textit{e.g.}][]{Wittkowski2004,Neilson2008}.  Observing the unresolved star's spectral energy distribution, it 
is possible to determine the bolometric flux, $F_\mathrm{bol}$, which 
depends on the star's angular diameter and effective temperature
\begin{equation} \label{eq:fbol}
F_{\rm{bol}} = 4\pi \theta^2_{\rm{Ross}} \sigma T_{\rm{eff}}^4,
\end{equation}
where $\theta_\mathrm{Ross}$ is the angular diameter at the Rosseland radius. 
Interferometry measures  spatial frequencies, from which one can infer the whole CLIV, hence a size scaling factor for any given limb-darkening profile. When combined with model stellar atmospheres, it is possible to determine the angular diameter where the intensity approaches zero and the optical depth approaches zero ($\theta_{\tau \rightarrow 0}$) which is designated $\theta_\mathrm{LD}$ 
\citep[\textit{e.g.}][]{Lester2013}.  As \citet{Wittkowski2004} have 
shown, $\theta_\mathrm{LD}$ is different from $\theta_\mathrm{Ross}$.  The two angular diameters are not independent variables but they can be independently measured using different methods. 

Rewriting Equation~\ref{eq:dr_r2} as a function of angular diameters rather than linear radii leads to the expression:
\begin{equation}\label{eq:dr_r4}
M_\star \propto T_{\rm{eff}} \frac{\theta_{\rm{Ross}}^2 d/2}{\theta_{\rm{LD}} - \theta_{\rm{Ross}}},
\end{equation}
where $d$ is the distance to the star and we replace $R_{\rm{Ross}} = d\theta_{\rm{Ross}}/2$ . 
This indicates the SMI can be determined from angular diameters derived from a combination of spectrophotometric and interferometric observations.  Testing the correlation 
between the relative atmospheric extension $\Delta R/R_\mathrm{Ross}$,
or its observed proxy 
$(\theta_\mathrm{LD} - \theta_\mathrm{Ross})/\theta_\mathrm{Ross}$, and 
$T_\mathrm{eff} (R_\mathrm{Ross}/M_\star)$ is the primary goal of this 
article.

To summarize this method, one can measure stellar masses using this relation plus interferometric and spectrophotometric observations in the following steps:
\begin{enumerate}
\item[i)] from spectrophotometric observations measure the combination of the effective temperature and Rosseland angular diameter;
\item[ii)] use a grid of model stellar atmospheres to fit interferometric observations to measure the limb-darkened angular diameter;
\item[iii)] take the combination of the Rosseland and limb-darkened angular diameters to measure $R_{\rm{Ross}}/M_\star$;
\item[iv]  and measure the radius of the star using an independent distance measurement plus the Rosseland angular diameter and then measure the stellar mass from the value of $R_{\rm{Ross}}/M_\star$.
\end{enumerate}
There are other paths to measure the stellar mass using the SMI correlation.  For instance, one can simultaneously fit the Rosseland angular diameter and effective temperature from spectrophotometry, the limb-darkened angular diameter from interferometric observations and apply our correlation as a constraint for the measurements to build a statistical fit of the fundamental stellar properties using grids of model stellar atmospheres.  Alternatively, one can assume an estimate for stellar properties, compute a single model stellar atmosphere to fit the interferometric observations and then measure a new value for mass using our correlation. From that measurement, we can compute a new model stellar atmosphere and repeat the process until the predicted values of mass converge.

We note that  our method takes advantage of the physics of model stellar atmospheres to refine measurements of stellar parameters in an already model-dependent analysis. All angular diameter measurements from interferometric observations to this point have assumed a model for stellar limb darkening, be it parametric, uniform disk, or a more-realistic model stellar atmosphere. 

 In terms of the intensity profile and fitting interferometric observations, we should ask what is the radius we measure.  Because interferometric observations are fit assuming some model of the intensity profile then the corresponding best-fit angular diameter is a function of that intensity profile in the limit of $\mu \rightarrow 0$.  When one assumes a uniform-disk intensity profile, one measures an angular diameter that is scaled by the flux.  When one assumes a plane-parallel model stellar atmosphere CLIV, then it is possible to measure a value of the Rosseland angular diameter, but only because plane-parallel atmospheres have no information of radius by definition.  The CLIV for spherically symmetric model atmospheres is far more complex and contains information of radius. The top of the photosphere, $R_{\rm{LD}}$, corresponds to the point where $\mu = 0$, hence the measured angular diameter will correspond to the $R_{\rm{LD}}$.  But, because the interferometric visibilities are the Hankel transform of the CLIV then the observations will be strongly weighted by the point where the CLIV drops off most rapidly.  That layer is approximately the Rosseland radius.  

Because of these issues, any attempts to measure the limb-darkened angular diameter requires interferometric visibilities that probe at least the second lobe of the visibility curve where differences in limb darkening at the edge of the stellar disk offer the greatest weight in the Hankel transform.  This implies that our proposed method is most useful for interferometric observations that reach baselines beyond the first minimum. 

Furthermore, one might expect that fitting interometric observations using model stellar atmospheres will allow one to uniquely measure the stellar mass. \cite{Neilson2008} found that the best-fit limb-darkening angular diameter from interferometric observations do not yield unique measurements of stellar properties.  \cite{Lester2013} presented a study of CLIV and spectra as a function of stellar mass.  If one knows precisely the stellar radius and effective temperature, then the stellar mass can be, at best, measured to a precision of a factor of a few.  Therefore, the combination of interferometric and spectrophotometric (or spectroscopic) observations is insufficient to measure stellar masses with any meaningful precision; our SMI correlation is crucial to improve those measurements.

\section{Model Stellar Atmospheres}\label{sec:sec3}
To test for the correlations suggested, we use a grid of model stellar atmospheres computed using the \textsc{SAtlas} code \citep{Lester2008, Neilson2013a}. These models are hydrostatic, spherically symmetric and use opacity distribution functions constructed from approximately two million atomic/ionic lines and eight million molecular lines extending to temperatures down to 2000~K  \citep{Castelli2010, Kurucz2011}. Convection is treated using the standard mixing length theory \citep{Vitense1953, Vitense1958}. This code is built upon the earlier \textsc{Atlas} codes developed by \cite{Kurucz1970}.  The \textsc{SAtlas} code computes intensities and fluxes as a function of wavelength assuming local thermodynamic equilibrium that can be applied to spectroscopic, photometric and interferometric observations.

 One caveat regarding the code is the omission of the extended 
molecular layers or MOLsphere \citep{Tsuji2000} of red supergiant stars. 
However, it should be noted that for many wavelengths the MOLsphere does 
not affect either the calculations or observations.  For example, 
observations of Betelgeuse in the $H$-band  \citep{Haubois2009}, a spectral 
band where the outer molecular layers would be expected to make a larger
contribution, did not detect the MOLsphere.  Furthermore, as of yet no 
model stellar atmosphere codes properly model the MOLsphere and interferometric observations of stars with MOLspheres can be interpreted in terms of multicomponent models such as a model stellar atmosphere CLIV plus an analytic MOLsphere component \citep{Montarges2014}.  That type of model limits bias in the measurement of $\theta_{\rm{LD}}$.  Similar biases, such as star spots and rotation, can also impact our method, but it is up to the observer to carefully interpret interferometric observations. 

\begin{figure}[t]
\begin{center}
\epsscale{1.2}
\plotone{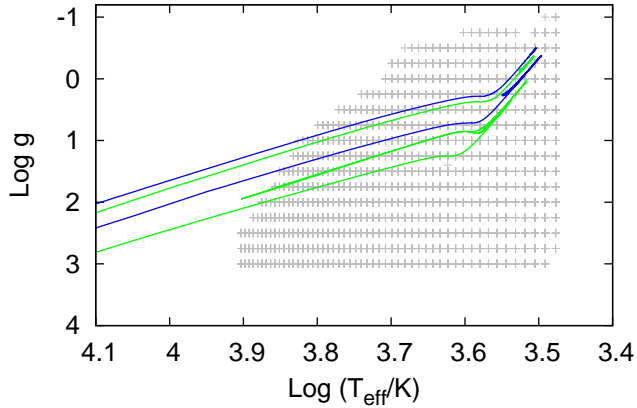}
\end{center}
\vspace{-0.8cm}
\caption{Spectroscopic HR diagram denoting the parameter space in $T_{\rm{eff}}$ and $\log g$ of the stellar atmosphere models considered in this work. The plus symbols  represent the 
$T_\mathrm{eff}$ and $\log g$ of the individual model atmospheres used in
this study.   The solid lines are post main-sequence evolutionary 
tracks for stars with masses 8 to $20~M_\odot$ from  \cite{Neilson2012b, Neilson2012c}.}\label{f1}
\end{figure}

The grid of spherical models used here spans $T_{\rm{eff}} = 3000~$to~$8000~$K in steps of 100~K, $\log g = -1$~to~$+3$ in steps of 0.25~dex and masses $M_\star = 0.5, 1, 2.5, 5, 7.5, 10, 12.5,  15, 17.5, 20~M_\odot$, although the parameter space has some gaps where models  failed to converge.  This grid of models was used by \cite{Neilson2013a} to compute limb-darkening coefficients, angular diameter corrections for interferometric observations and gravity-darkening coefficients. Figure~\ref{f1} plots the model parameter space on a 
spectroscopic Hertzsprung-Russell diagram \citep{Langer2014} along with 
a sample of stellar evolution tracks for masses from 8 to $20~M_\sun$ 
\citep{Neilson2012b, Neilson2012c}.

For each spherical model stellar atmosphere in the grid, the intensity is computed as a function of $\mu$, which is defined as the cosine of the angle formed between the  normal vector at a point on the stellar surface and the direction toward the distant observer.  The intensity was computed for 1,000 equally spaced points in $\mu$-space, making these the most detailed intensity profiles published \citep{Neilson2013a,Neilson2013b}.  

\section{Results} \label{sec:sec4}
For each spherical model stellar atmosphere in the grid we compute the surface radius $R_{\tau \rightarrow 0}$, which we 
consider to be the limb-darkened radius. In addition to  this limb-darkening radius and the Rosseland radius $R_{\tau = 2/3}$, we also have the 
uniform-disk radii for interferometric observations from \cite{Neilson2013a}, which is defined as the stellar radius necessary to fit interferometric visibilities assuming the CLIV is constant across the stellar disk. It should be noted that the calculations by \cite{Neilson2013a} predicted the relative uniform-disk correction, i.e., $R_{\rm{UD}}/R_{\rm{LD}}$ or $R_{\rm{UD}}/R_{\rm{Ross}}$ and not the absolute uniform-disk values.  

This correlation is computed for \textsc{SAtlas} stellar atmosphere models and is based on the parameters describing each model.  In particular, \textsc{SAtlas} defines the limb-darkened radius where $\tau_{\rm{Ross}}\rightarrow 0$.  Slightly different assumptions and implementations used in other codes, such as \textsc{Phoenix} \citep{Hauschildt1999} and 
\textsc{MARCS} \citep{Gustafsson2008}, might yield slightly different
results.

 The value of the limb-darkened radius and $\theta_\mathrm{LD}$ 
depends on the definition adopted for the star's surface as the limit where $\tau_{\rm{Ross}} \rightarrow 0$.  For example, 
the \textsc{Atlas/SAtlas} codes retain the original \citet{Kurucz1970} location of the top of 
the atmosphere at $\tau_\mathrm{Ross} = 1.334 \times 10^{-6} \
(\log_{10} \tau_\mathrm{Ross} = -5.875)$.  This choice 
of the surface $\tau_\mathrm{Ross}$, however, has essentially no effect on the 
physical radius because the gas density is so low.  As an example, for a 
typical red giant model atmosphere with 
$L_\star = 500 \ L_\sun, \ M_\star = 1 \ M_\sun$ and 
$R_\star = 50 \ R_\sun$, redefining the physical radius to the location where the 
Rosseland optical depth is $100 \times$ larger than the 
value adopted here reduces the radius by 
less than $0.1\times$ our uncertainty of ($\theta_{\rm{LD}} -\theta_{\rm{Ross}}$), but has no effect on the fit to interferometric data.  The definition of $\theta_\mathrm{LD}$ is a convention that depends on the CLIV.  As long as the same definition is used by both stellar atmosphere models and by the model fit to interferometric data, $\theta_{\rm{LD}}$ acts strictly as a spatial scaling parameter for the CLIV profile.  Thus it possesses a unique, non-ambiguous nominal value.  We note, however, that the $\theta_{\rm{LD}}$ conventions will  be equivalent to $\theta_{\rm{Ross}}$ only in the limit where the atmospheric extension goes to zero.  We conclude 
that our definition of the limb-darkened radius, although model 
dependent, is still a useful tool for interferometric observations, 
especially when the model intensity profiles are used to fit those 
observations.

\subsection{Correlations}\label{sec:sec4.1}

\begin{figure*}[t]
\begin{center}
\epsscale{1.1}
\plottwo{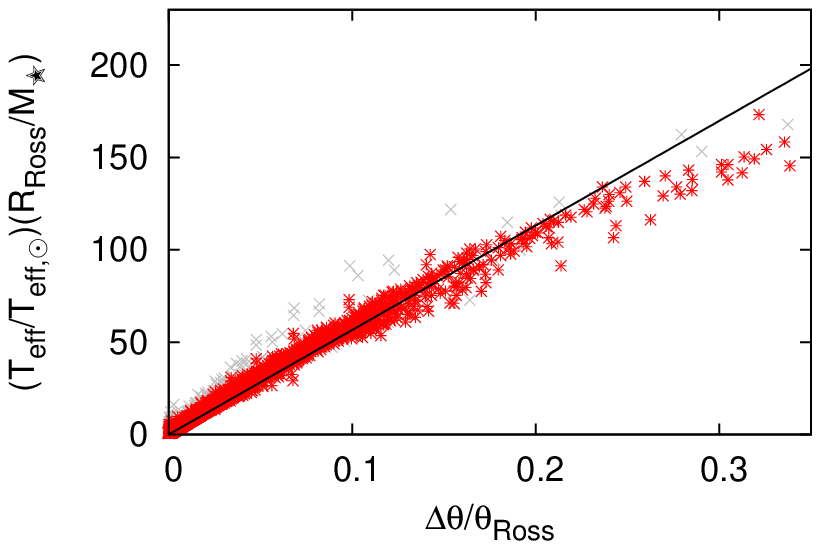}{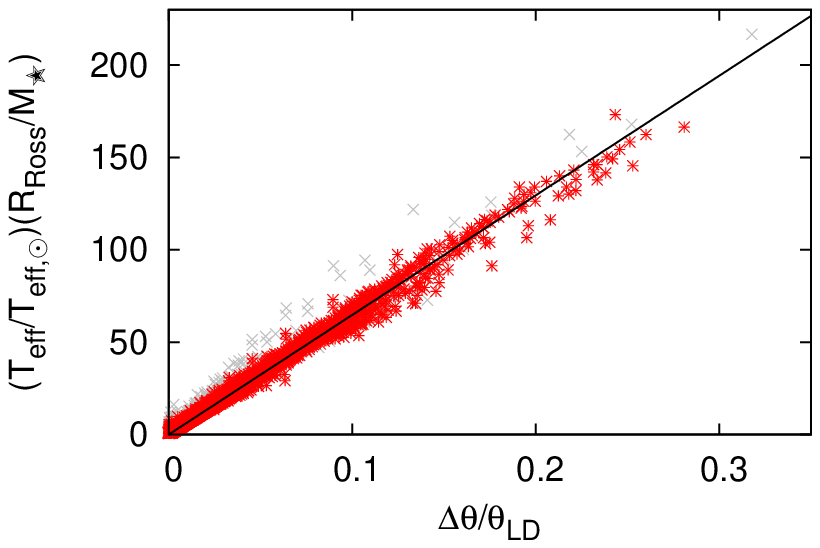}
\end{center}
\vspace{-0.8cm}
\caption{The correlation between the product of $T_{\rm{eff}} \times$ SMI and the two proxies for the atmospheric extension. (Left) $(\theta_\mathrm{LD} - \theta_\mathrm{Ross})/\theta_\mathrm{Ross}$.  There is a  correlation between the variables suggesting the potential for measuring fundamental stellar parameters from measurements of angular diameters, however, for large values of $(\theta_\mathrm{LD} - \theta_\mathrm{Ross})/\theta_\mathrm{Ross}$ the models seem to follow a non-linear trend. (Right) An even stronger correlation is seen for the product of $T_{\rm{eff}} \times$ SMI and $(\theta_\mathrm{LD} - \theta_\mathrm{Ross})/\theta_\mathrm{LD}$. }\label{fc1}
\end{figure*}

For the first test, we search for correlations between the atmospheric extension  $(\theta_\mathrm{LD} - \theta_\mathrm{Ross})/\theta_\mathrm{Ross}$ and a measure of $T_{\rm{eff}} \times$ SMI. In Figure~\ref{fc1}, we plot the two measurements and confirm the existence of a linear correlation.  Models with $T_{\rm{eff}} \le 3200~$K are shown in gray, but they are not used in fitting the correlations because at low temperature the model stellar atmospheres do not converge as well, implying a significant flux error that will bias any fit. The best-fit relation is
\begin{equation}\label{eq:ross}
\left(\frac{T_{\rm{eff}}}{T_{\rm{eff},\odot}}\right)\left(\frac{R_{\rm{Ross}}}{M_\star}\right) = (565.99 \pm 0.96) \frac{\theta_\mathrm{LD} - \theta_\mathrm{Ross}}{\theta_\mathrm{Ross}}.
\end{equation}

This is a strong correlation, confirming our expectations from the previous sections. However, we note that the data points for $(\frac{T_\mathrm{eff}}{T_{\mathrm{eff},\sun}})(\frac{R_\mathrm{Ross}}{M_\star}) > 120 \ \frac{R_\sun}{M_\sun}$ drift above the best-fit straight line, which is dominated by the large 
number of points $\leq 120 \ \frac{R_\sun}{M_\sun}$.  This drift could be due to a change in the dominant opacity for the most extended models, which are also the stars with the lowest effective temperatures.  This 
deviation deserves further investigation, but for the purposes of this work we consider only a linear fit.

Because of the deviation in the plots in the left panel of Figure~\ref{fc1}, we also fit a slightly different correlation, $\Delta \theta/\theta_\mathrm{LD}$, instead of normalizing with respect to $\theta_\mathrm{Ross}$.  There is no obvious justification for this revised relation, but the correlation is stronger, particularly at large atmospheric extensions.
\begin{equation}\label{eq:ld}
\left(\frac{T_{\rm{eff}}}{T_{\rm{eff},\odot}}\right)\left(\frac{R_{\rm{Ross}}}{M_\star}\right) = (647.48\pm 0.67)\frac{\theta_\mathrm{LD} - \theta_\mathrm{Ross}}{\theta_\mathrm{LD}}.
\end{equation}
 A possible reason for this improvement is that at the largest extensions $T_{\rm{eff}}$ is greater, but the gravity is weaker. The radiation pressure is increased relative to the gas pressure, which makes it somewhat more numerically difficult to compute stable stellar atmospheres. This may result in an overestimation of the limb-darkened angular diameter, and thus in deviations of the parameter $\theta_{\rm{LD}}$ at the largest extension, that the reformulated relation would then partially cancel out.

The second test is to measure the uniform-disk angular diameters from interferometric observations \citep[\textit{e.g.}][]{Davis2000, Belle2009} and lunar occultation observations \citep[\textit{e.g.}][]{Richichi2014} and other methods. This is important because the more distant a star, the harder it is to resolve its second interferometric lobe. But one may still be able to measure the uniform-disc angular diameter in different wavelengths, and these have been tabulated over the years for various types of stars \citep[\textit{e.g.}][]{Belle2009,Boyajian2014, vonBraun2014}.  We test various combinations of the uniform-disk, Rosseland and limb-darkened angular diameters for correlations with the atmospheric extension. While the Rosseland and limb-darkening angular diameters are independent of wavelength in the models, the uniform-disk radii are not. The limb-darkened angular diameters in the models are defined by the limit as $\tau_{\rm{Ross}} \rightarrow 0$ and $\mu \rightarrow 0$.

\begin{figure*}[t]
\begin{center}
\epsscale{1.1}
\plottwo{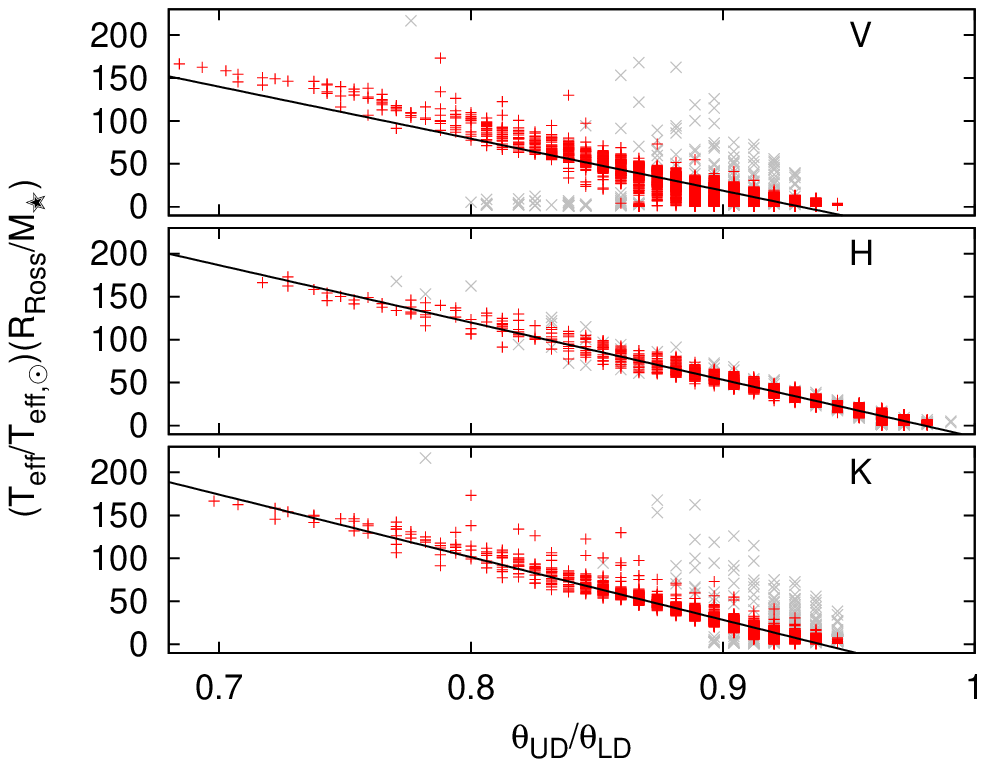}{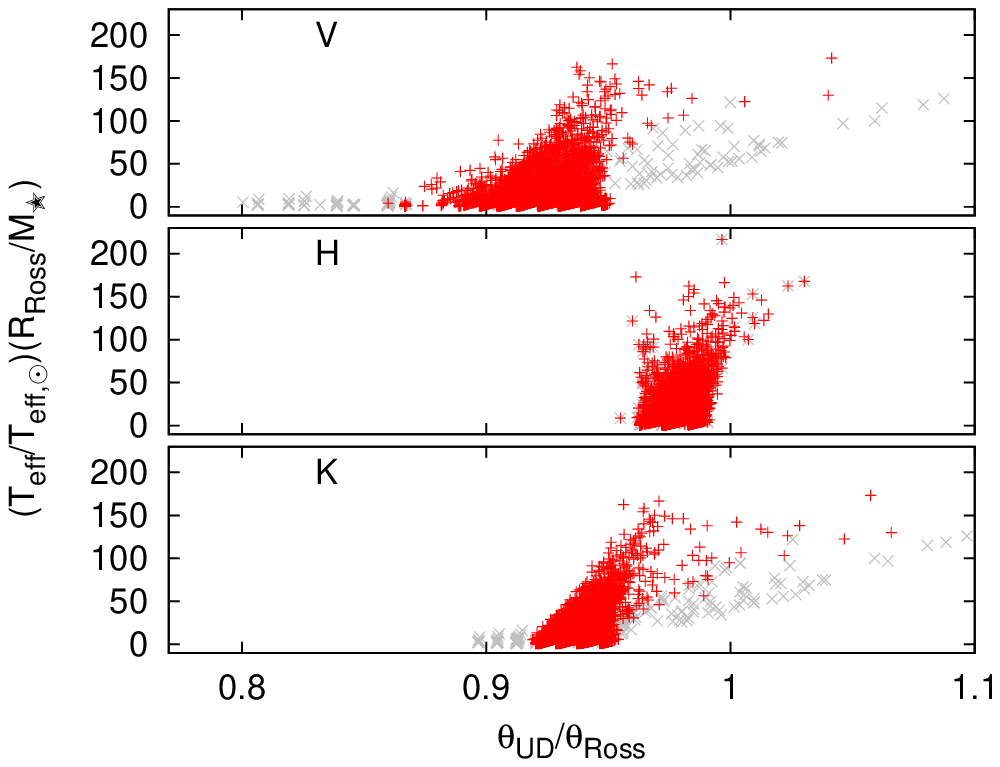}
\end{center}
\vspace{-0.6cm}
\caption{Correlation between the $T_{\rm{eff}}R_{\rm{Ross}}/M_\star$ and the proxies for the atmospheric extension, (left) $\theta_{\rm{UD}}/\theta_{\rm{LD}}$,  (right) $\theta_{\rm{UD}}/\theta_{\rm{Ross}}$  for different wavebands: $V, H$, and $K$-band.  Correlations are weaker when the uniform-disk angular diameters are considered, but trends are apparent for the ratio of $\theta_{\rm{UD}}/\theta_{\rm{LD}}$.  There is no obvious correlation between $T_{\rm{eff}}R_{\rm{Ross}}/M_\star$ and the ratio $\theta_{\rm{UD}}/\theta_{\rm{Ross}}$ that suggests this is the limiting limb-darkening assumption for our method.}\label{io}
\end{figure*}

In Figure~\ref{io} we plot the uniform-disk angular diameters computed from synthetic visibilities relative to the limb-darkened angular diameters (left) and Rossland angular diameters (right) as functions of atmospheric extension.  The correlations show greater scatter, but are still sufficiently strong to 
indicate that replacing $\theta_\mathrm{Ross}$ by $\theta_\mathrm{UD}$ in 
the atmospheric extension term defined in Equation~\ref{eq:dr_r4} and shown in Figure~\ref{fc1} would still exhibit a significant correlation, even more so in $H$-band, indicating further exploration is warranted.  Unfortunately, there appears to be no correlation between the SMI and the ratio $\theta_{\rm{UD}}/\theta_{\rm{Ross}}$, which indicates that for the worst possible case the correlations break down.

\subsection{Testing Atmospheric Extension Correlations}

As a consistency check, we use the results for $\chi$~Phe from \cite{Wittkowski2004}, who measured an effective temperature $T_{\rm{eff}} = 3550\pm 50$~K and a Rosseland angular diameter  $\theta_{\rm{Ross}} = 8.0\pm0.4$~mas solely from spectrophotometric observations.  \cite{Wittkowski2004} also derived a limb-darkened angular diameter of $8.65 \pm 0.15$~mas by fitting \textsc{Phoenix} model stellar atmospheres to optical interferometric observations.  These fits use an entirely different model atmosphere code than is used in our fits, but similar results can be found using \textsc{SAtlas} models \citep{Neilson2008}.  The values found by \cite{Wittkowski2004} give $\Delta \theta /\theta_{\rm{Ross}} = \Delta R/R_{\rm{Ross}} =  0.081\pm0.005$. Using this and the $T_\mathrm{eff}$ derived by 
\citet{Wittkowski2004} in our relation given in Equation~\ref{eq:ross} yields the SMI  $= 74.5\pm 5.7$ in solar units.  \cite{Wittkowski2004} derived the radius to be $85\pm 10~R_\odot$.  When we combine that radius with the SMI we get a stellar mass $M_\star= 1.14\pm 0.22~M_\odot$. This is consistent with the value found by \cite{Wittkowski2004}, but it uses a less demanding measurement of the Rosseland angular diameter and does not invoke stellar evolution models.  If we use Equation~\ref{eq:ld} instead, we measure a mass $M_\star=1.07\pm0.21~M_\odot$, which is slightly smaller, but consistent with our first estimate and the previous reports.

 Assuming the best method is to measure $\theta_{\rm{Ross}}$ from spectrophotometric observations, then we also have a robust measure of the effective temperature, which suggests using Equation~\ref{eq:ross} to find the SMI.  To extract the stellar mass from the SMI we seem to require some independent measure of the radius, say by using the angular diameter and a measurement of the star's parallax.  However, we do not necessarily require a measurement of the distance. If we have spectroscopic observations that allow for a fit of effective temperature and gravity, then we can combine the gravity and the SMI to extract both the Rosseland radius and the mass, and one also can use this method to find the distance to the star.  To conclude, there are a number of potential methods to measure the stellar mass from our atmospheric extension, independent of observations of binary stars or asteroseismic relations \citep{Huber2010, Kallinger2010, Stello2013}.
 

\section{Summary}
The goal of this work was to develop a new method for inferring stellar masses from measurements of atmospheric extension in stars. We use the previously published grid of \textsc{SAtlas} model stellar atmospheres \citep{Neilson2013a} and compute values of the Rosseland, limb-darkened and uniform-disk radii, where combinations of any two parameters are proxies for the atmospheric extension.  We also show that another proxy for the atmospheric extension is $T_{\rm{eff}}R_{\rm{Ross}}/M_\star$, which leads us to define stellar mass index, SMI $\equiv R_{\rm{Ross}}/M_\star$. 

We develop a methodology for measuring stellar masses from the trio of model stellar atmospheres, interferometric observations and spectrophotometic observations. That method takes independent measurements of  Rossland and limb-darkened angular diameters and computes the stellar mass index, $R_{\rm{Ross}}/M_\star$, from the combination of the two diameters. With an independent measurement of the distance or the stellar gravity, then the mass can be measured without invoking stellar evolution tracks.

We then test these different proxies for correlations and find that the combination of  Rosseland and limb-darkened radii are strongly correlated with the atmospheric extension, which provides a powerful method for inferring stellar parameters from interferometric and spectrophotometric observations.  In addition, we test for correlations with other combinations, including the more easily measured uniform-disk radius, but find no significant correlation in that case.

The method outlined here builds upon the technique described by 
\citet{Neilson2012}, but it does not employ a specific limb-darkening 
law with ad-hoc coefficients that can be sensitive to the quality of the observations 
\citep[\textit{e.g.}][]{Dominik2004}.  By focusing on angular diameters, 
one can employ more readily available observations from a plethora of 
sources, such as interferometry and lunar occultations, that can be 
coupled with spectrophotometric and/or spectroscopic observations.
 Our preliminary test of the method is based on the interferometric data of \citet{Wittkowski2004}. It demonstrates that spherical atmosphere models, plus the correlations derived here, provide realistic fits to the radii and masses of giant stars and presumably supergiant stars as well.

 We chose here to consider stars with stellar extensions large enough that the ratio $(\theta_\mathrm{LD} - \theta_\mathrm{Ross})/\theta_\mathrm{Ross}$ will have small uncertainties when measured by optical interferometry, but the method is not strictly limited to evolved or cool stars. To further validate it, interferometric observations will need to be carried out on stars that can be reliably modelled by stellar atmosphere codes. In particular, in the near future we plan on observing eclipsing double-line binaries for which the masses and radii of the stars are known.

\acknowledgements
We wish to thank the referee for comments that significantly improved this work. FB acknowledges funding from NSF award AST-1445935.
\bibliographystyle{aa} 

\bibliography{ld} 

\end{document}